\documentclass[acmsmall,screen,review=false]{acmart}
\usepackage{balance}
\usepackage[utf8]{inputenc}
\usepackage{graphicx}
\date{August 2021}
\usepackage{float}
\usepackage{array}
\usepackage{tabularx}
\usepackage{multirow}
\usepackage{amsmath}
\usepackage{amssymb}
\usepackage{threeparttable}
\usepackage{setspace}
\usepackage{color}
\usepackage{colortbl}
\usepackage{xcolor}
\usepackage{soul}
\usepackage{booktabs}
\usepackage{makecell}
\usepackage{footnote}
\usepackage{algorithm} 
\usepackage{algorithmicx}
\usepackage{algpseudocode}
\usepackage{hyperref}
\usepackage{caption}
\usepackage{verbatim}
\usepackage{ulem}
\usepackage{lineno,hyperref}

\definecolor{mygray}{gray}{.9}
\definecolor{mypink}{rgb}{.99,.91,.95}
\definecolor{mycyan}{cmyk}{.3,0,0,0}
\definecolor{myyellow}{RGB}{255,230,204}
\definecolor{mybule}{RGB}{218,232,252}
\definecolor{mygreen}{RGB}{213,232,212}
\definecolor{titleColor}{RGB}{102,102,102}

\renewcommand\footnotetextcopyrightpermission[1]{} 
\title{API Entity and Relation Joint Extraction from Text via Dynamic Prompt-tuned Language Model}
\begin{document}


\author{Qing~Huang}
\affiliation{%
  \institution{Jiangxi Normal University, School of Computer Information Engineering}
  \city{Nanchang}
  \state{Jiangxi}
  \country{China}}
\email{qh@whu.edu.cn}

\author{Yanbang~Sun}
\authornote{Y. Sun and Q. Huang are co-first authors.}
\affiliation{  
  \institution{Jiangxi Normal University, School of Computer Information Engineering}
  \city{Nanchang}
  \state{Jiangxi}
  \country{China}}
\email{ybsun@jxnu.edu.cn}

\author{Zhenchang~Xing}
\affiliation{\institution{CSIRO's Data61 \& Australian National University, College of Engineering and Computer Science}
  \city{Canberra}
  \country{Australia}}
\email{zhenchang.xing@data61.csiro.au}

\author{Min~Yu}
\authornote{M. Yu is the corresponding author.}
\affiliation{\institution{Jiangxi Normal University, School of Computer Information Engineering}
  \city{Nanchang}
  \state{Jiangxi}
  \country{China}}
\email{myu@jxnu.edu.cn}

\author{Xiwei Xu}
\affiliation{\institution{CSIRO's Data61}
  \city{Sydney}
  \country{Australia}}
\email{xiwei.xu@data61.csiro.au}

\author{Qinghua~Lu}
\affiliation{\institution{CSIRO's Data61}
  \city{Sydney}
  \country{Australia}
}
\email{qinghua.lu@data61.csiro.au}

\begin{abstract}
Extraction of Application Programming Interfaces (APIs) and their semantic relations from unstructured text (e.g., Stack Overflow) is a fundamental work for software engineering tasks (e.g., API recommendation).
However, existing approaches are rule-based and sequence-labeling based.
They must manually enumerate the rules or label data for a wide range of sentence patterns, which involves a significant amount of labor overhead and is exacerbated by morphological and common-word ambiguity.
In contrast to matching or labeling API entities and relations, this paper formulates heterogeneous API extraction and API relation extraction task as a sequence-to-sequence generation task, and proposes AERJE, an API entity-relation joint extraction model based on the large pre-trained language model.
After training on a small number of ambiguous but correctly labeled data, AERJE builds a multi-task architecture that extracts API entities and relations from unstructured text using dynamic prompts.
We systematically evaluate AERJE on a set of long and ambiguous sentences from Stack Overflow.
The experimental results show that AERJE achieves high accuracy and discrimination ability in API entity-relation joint extraction, even with zero or few-shot fine-tuning.
\end{abstract}


\maketitle

\section{INTRODUCTION}\label{sec:sec1}
Application Programming Interfaces (APIs) are important software engineering artifacts that can be frequently found in a wide range of natural language texts, from official API references and tutorials to informal online forums.
Meanwhile, API relations are also embedded in these texts.
For example, the text ``To manipulate data you actually need \textit{executeUpdate()} rather than \textit{executeQuery()}'' in the Stack Overflow (SO) post~\footnote{\href{https://stackoverflow.com/questions/1905607}{https://stackoverflow.com/questions/1905607}} describes the Function-Replace relation~\cite{9894095} between \textit{executeUpdate()} and \textit{executeQuery()}.
This API relation reveals that we should replace \textit{executeQuery()} with \textit{executeUpdate()} to solve the question in the post, i.e., ``why cannot issue data manipulation statements with \textit{executeQuery()}''.
API entity and relation extraction from unstructured texts is fundamental for efficiently accessing and applying API knowledge to various software engineering tasks.
Once extracted, these entities and relations can be organized into structured knowledge (particularly in the form of knowledge graphs) to support a variety of software engineering tasks such as API linking~\cite{ye2018apireal, Dagenais2012RecoveringTL}, API recommendation~\cite{huang2018api,Xie2020APIMR}, and API comparison~\cite{liu2020generating}.

\begin{table}[t]
\caption{Three types of ambiguities for API entities and relations.}
\label{tab:table1_amb}
\begin{threeparttable}
\begin{tabular}{l|l}
\hline
PostID     & Sentence                                                                             \\ \hline
\#47871272 & You need to override \underline{remove()} in your iterator.                                      \\
\#14200489 & This code is invalid since \underline{l.remove()} is called during iteration over l.             \\
\#60017952 & You may be calling \underline{iterator.remove} more than once.                                      \\ \hline
\#34682267 & By default, \underline{printWriter} calls flush in \underline{println}, whereas it doesn't do this in \underline{print}. \\
\#703396   & If the idea is to \uwave{print} integer stored as doubles...     \\ \hline
\#322715   & \underline{linkedlist} and \underline{arraylist} are two different implementations of the \underline{list} interface.     \\
\#33405095  & \underline{nextline()} will read the entire line, but \underline{next()} will only read the next word.                                  \\
\#355089   & \underline{StringBuffer} is synchronized, \underline{StringBuilder} is not.                                  \\ \hline
\end{tabular}
\begin{tablenotes}
    \normalsize
    \item Note: API mention is tagged with an underline; common word is tagged with a wavy line.
\end{tablenotes}
\end{threeparttable}
\end{table}

There are currently two main types of approaches for extracting API entities from unstructured text.
The first is a rule-based approach such as language-convention based regular expressions~\cite{bacchelli2010linking,Treude2016AugmentingAD}, island parsing~\cite{Rigby2013DiscoveringEC,Bacchelli2011ExtractingSD} and heuristic rule matching~\cite{9894095,liu2020generating,Ren2020APIMisuseDD}.
Because it is impossible to manually enumerate the rules that adapt to all sentence patterns, it suffers from rule design overhead.
The second is a sequence labeling based approach such as CRF~\cite{ye2018apireal,Ye2016LearningTE} and Bi-LSTM-CRF~\cite{Huo2022ARCLINAA}.
Because it is impossible to manually label entities for a large amount of sentences, it suffers from data labeling overhead.
Compared with API entity extraction, relation extraction from software text is rather primitive, which relies on either API syntax (e.g., a class declares a method)~\cite{liu2020generating}, special-tag annotated relations (e.g., ``see also'' keyword and  hyperlink-based method)~\cite{Li2018ImprovingAC}, or some ad-hoc relation phrases (e.g., ``differ in'' and ``be similar to'')~\cite{9894095}.
Same as rule-based entity extraction, these relation extraction methods suffer from rule-design overhead.
We refer to rule design overhead and data labeling overhead as labor overhead in this work.

This labor overhead is exacerbated by three types of ambiguities, which necessitate the manual design of more rules or the labeling of more data to distinguish ambiguous sentences.
Morphological ambiguity, which includes abbreviations, synonyms, and misspellings, is one type of ambiguity~\cite{Ye2016LearningTE}.
It is common in informal discussions, because people rarely write full API names that exactly match the API names in the library~\cite{Chen2017UnsupervisedSM}.
Three sentences in the first row of Table~\ref{tab:table1_amb}, for example, shows three morphological variations of API \textit{java.util.iterator.remove()}, that either omit some prefixes and special symbols (e.g., package names, class names, and ``\textit{()}''), or are preceded by user-defined variable names.
The second type is common-word ambiguity between common words and API mentions~\cite{Ye2016LearningTE}, which occurs because people frequently write API method names without proper punctuation, parentheses, and uppercase letters.
For example, as shown in the second row of Table~\ref{tab:table1_amb}, even though the word \textit{print} appears in two sentences, \textit{print} in the first sentence refers to the API \textit{java.io.printwriter.print()}, whereas \textit{print} in the second one is only a verb.
The final type is expression ambiguity of API semantic relations, which is caused by changes in sentence patterns.
In general, the same API relation can be expressed in multiple sentence patterns.
For example, as shown in the third raw of Table~\ref{tab:table1_amb}, three sentence patterns are used in the three sentences, all of which express the Behavioral-Difference relation between API entities.
They are ``API1 and API2 are different'', ``API1 does one thing, but API2 does the other thing'', and ``API1 is (adjective), API2 is not''.

To alleviate the labor overhead, we devise a novel idea of extracting API entities and relations using a large pre-trained language model (LLM). 
LLM stores a large amount of prior knowledge and can serve as a neural knowledge base of real-world entities and relations~\cite{Petroni2019LanguageMA}.
In addition, LLM can provide better model initialization~\cite{Qiu2020PretrainedMF} and strong learning capbility.
By fine-tuning a LLM with a small set of domain-specific training data,  we can prompote the LLM to identify as many API entities and relations as possible. 
In order to make LLM be more discriminative, the training data should contain sufficient morphological and common-word ambiguity, and the API entities and relations should be labeled correctly. 
To reduce manual labeling, we devise morphology and verb-based data augmentation strategies to generate more ambiguous data but correctly labeled sentences for the LLM fine-tuning.

Existing work~\cite{ma2019easy} separates API entity extraction and relation extraction as two tasks, leaving relation extraction heavily reliant on entity extraction results, which leads to error propagation~\cite{Zhong2021AFE}.
Instead, we consider API entity extraction and relation extraction as correlated tasks and adopt a unified model for joint entity and relation extraction, inspired by the recent work on universal information extraction (UIE~\cite{Lu2022UnifiedSG}).
However, the LLM (i.e., T5~\cite{Raffel2020ExploringTL}) in UIE often fails with complex sentences, particularly long and ambiguous sentences containing API entities and various relations, because it only uses one static prompt to recognize all types of API relations.
To tackle this issue, we design a dynamic prompt generator, inspired by the input-dependent prompt tuning method~\cite{Levine2022StandingOT}, that generates dynamic prompts for a small number of potentially relevant relations at inference time based on the actual input sentences, rather than relying on the same static prompt for all inputs.
When confronted with complex sentences, the more relation types to recognize, the more noise it suffers from, the more difficult it is for LLM to understand to what extent a complex sentence contains certain relations.
As our dynamic prompt reduces the number of relation types to recognize and mitigate noise interference, it improves the extraction accuracy of API relations.

In this paper, we propose a API Entity-Relation Joint Extraction framework, called AERJE.
It consists of a dynamic prompt generator and a joint entity-relation extractor.
The kernel of the prompt generator is a BERT-based text classifier that is used to classify the input text.
Each class represents an API relation and the prompt generator generates dynamic prompts based on the top-N possible API relations.
The generated dynamic prompt and the input sentence are fed into the joint entity-relation extractor to extract the API entities and relations contained in the text.
In our current implementation, the joint entity-relation extractor is Transformer-base LLM (T5).
The prompt generator and the entity-relation extractor are fine-tuned in an end-to-end manner.

No model, to the best of our knowledge, can simultaneously extract both API entities and relations.
AERJE, on the other hand, achieves an F1 score of 96.51\% for API entity extraction, which is approximately 6\% higher than the state-of-the-art API entity recognition model ARCLIN~\cite{Huo2022ARCLINAA} and 7\% higher than APIReal~\cite{ye2018apireal}, and an F1 score of 81.20\% for API relation extraction.
Then, we evaluate the impact of intrinsic factors (two data augmentation strategies and the number of API relations in the dynamic prompts) on performance.
Our experiments find that data augmentation helps to improve AERJE's discriminative capability for API entities and relations, and the dynamic prompts with four API relations can significantly improve AERJE's extraction accuracy.
Finally, we assess AERJE's generalization and ability to extract API entities and relations in low-resource scenarios (i.e., less than 0.8\% fine-tuning data) and find that, even under low resource conditions, our AERJE still has strong extraction ability, outperforming APIReal~\cite{ye2018apireal} and ARCLIN~\cite{Huo2022ARCLINAA}.

The main contributions of this paper are as follows:
\begin{itemize} 
    \item 
    Conceptually, we are the first to formulate heterogeneous API extraction and API relation extraction tasks as a uniform sequence-to-sequence generation task, and propose AERJE, an API entity-relation joint extraction framework based on pre-trained LLMs.
    \item
    We devise two data augmentation strategies in order to obtain more ambiguous but correctly labeled sentences. Learning such sentences enables AERJE to be more discriminative for API entities and relations.
    \item
    Unlike the single task model, we build a multi-task architecture that encodes the structures of entity and relation extraction into a unified structure language for extracting API entities and relations simultaneously.
    \item
    Instead of using a single static prompt with all types of API relations for all sentences, we design a dynamic prompt based on relation classification, which reduces the number of relation types to recognize, eliminates noise interference, and lowers the difficulty of relation extraction.
    \item
    We systematically evaluate the AERJE's intrinsic factors, performance, generalization, and few-shot learning capabilities.
    It is the first approach to extract API entities and relations simultaneously, and it achieves superior performance than independent API extraction and API relation extraction.
    Our data package can be found here\footnote{\href{https://anonymous.4open.science/r/AERJE-6DBF/README.md}{https://anonymous.4open.science/r/AERJE-6DBF/README.md}}, the code will be released after the paper is accepted.
\end{itemize}
\section{APPROACH}
\begin{figure*}[t]
    \centering
    \includegraphics[width=1\textwidth]{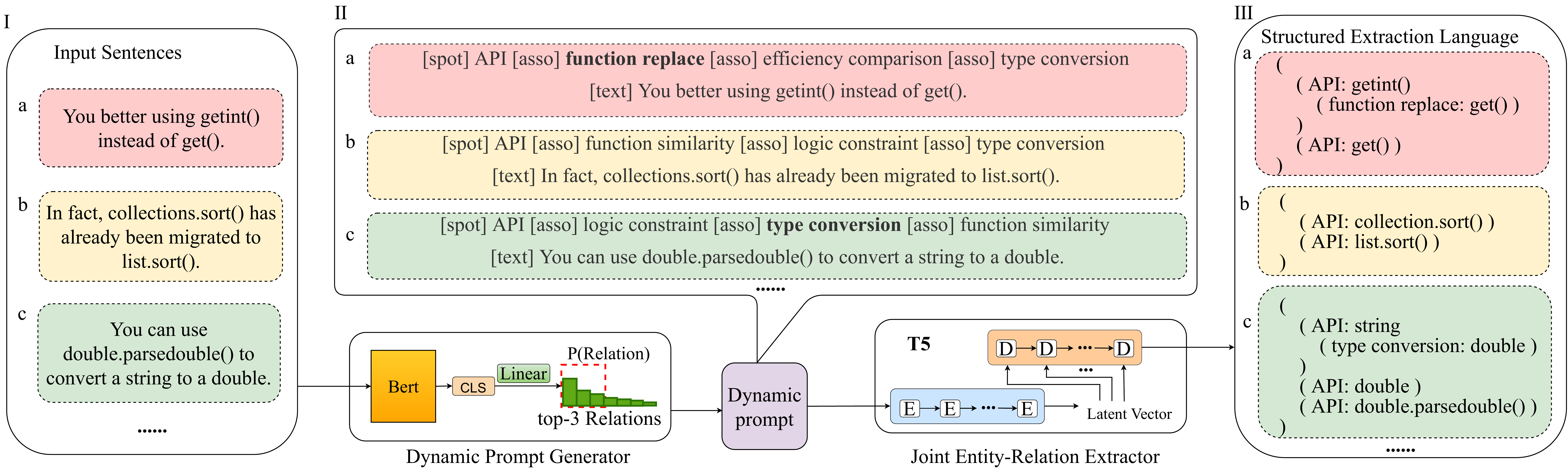}
    \caption{Overall Framework of AERJE.
    The Dynamic prompt's bold font represents the semantic relation to be extracted from the input sentence. 
    II.b lacks a bold font because I.b contains no semantic relation.}
    \label{fig:frame_fig}
\end{figure*}

We formulate heterogeneous API extraction and API relation extraction tasks as a uniform sequence-to-sequence generation task, and propose a novel model AERJE to accomplish it.
As shown in Fig.~\ref{fig:frame_fig}, AERJE consists of a dynamic prompt generator and a joint entity-relation extractor.
The dynamic prompt generator generates dynamic prompts based on the input texts (one at a time).
The input text is then appended to the prompt to form a whole input that is fed into the joint entity-relation extractor, which generates a structured extraction language sequence with API entities and relations.
\subsection{Dynamic Prompt Generator}
This section describes how to build a prompt that unifies heterogeneous API extraction and API relation extraction tasks, followed by a discussion of how to design dynamic prompt to improve AERJE's API relation extraction performance.
\subsubsection{Prompt Construction for Multi-tasking}\label{sec:procon}
In order to extract both API entities and API relations from an input text, the prompt consists of API entity type, API relation type, and the input text, which are labeled by [spot], [asso] and [text], respectively.
For example, ``[spot] API [asso] function replace [asso] efficiency comparison [text] You better using getint() instead of get()'' represents an API entity type ``API'', two relation types ``function replace'' and ``efficiency comparison'', and an input text ``You better using getint() instead of get()''.
In this work, we consider a generic API entity type ``API'' and seven relation types defined in~\cite{9894095}, including ``function similarity'', ``behavior difference'', ``logic constraint'', ``type conversion'', ``function collaboration'', ``efficiency comparison'', ``function replace''.
Note that more fine-grained API entity types can be used, such as ``class'', ``method'', ``field''~\cite{ye2016software}, but we leave it as our future work.
\subsubsection{Dynamic Prompt Generation}\label{sec:dypro}
As stated in Section~\ref{sec:sec1}, the more relation types there are, the harder it is for T5 to determine which types of relation the API entities in the input text belong to,  especially when the sentence is long and ambiguous.
If we adopt the static prompt that includes all seven relation types, the relation extraction performance of the model will decrease (cf. RQ3).
As a result, we design a dynamic prompt generation method to make the content of the prompt more accurate and instructive for the complex input text.
The dynamic prompts, as shown in II.a of Fig.\ref{fig:frame_fig}, contain only the top-N relations and provide better guidance to the subsequent T5-supported joint entity-relation extractor.
Here, the prompt generator is implemented as a text classifier which predicts the API relations present in the input text.
We use a BERT-based classifier because the pre-training task (i.e., Next Sentence Prediction) of BERT~\cite{Devlin2019BERTPO} is consistent with our task, both of which are classification tasks.
Given a sentence containing API entities (see I.a of Fig.~\ref{fig:frame_fig}), the BERT-based classifier outputs the probability that the sentence belongs to each semantic relation; the top-3 relations are then chosen as candidate relations.
Finally, entity type, candidate relations, and input sentence are connected by labels (i.e., [spot], [asso], [text]) to generate the dynamic prompt (see II.a of Fig.~\ref{fig:frame_fig}).

Note that the BERT-based classifier in our current implementation aims to narrow the scope and provide candidate relations, and it cannot replace the API relations extractor.
When the candidate relations classified by the classifier do not fit these entities in the sentence, the extractor does not force a relation to be selected from the incorrect candidate relations, but instead assumes that no relation exists between these entities.
For example, given a sentence with no relations between API entities (see I.b of Fig.\ref{fig:frame_fig}), the dynamic prompt generator generates a dynamic prompt (see II.b of Fig.\ref{fig:frame_fig}).
Based on such a dynamic prompt, the subsequent extractor will not extract relations from the sentence as none of the candidate relation types is applicable to the input sentence.

To summarize, too many candidate relations may reduce the extractor's ability to recognize them, while too few candidate relations may cause the extractor to miss the correct relations.
As a result, we should investigate the appropriate number of candidate relations (cf. RQ3).

\subsection{API Joint Entity-Relation Extractor}
We adopt a structured extraction language (SEL)~\cite{Lu2022UnifiedSG} to encode the structures of entity extraction and relation extraction into a unified representation, so that heterogeneous API extraction and API relation extraction tasks can be modeled uniformly within a sequence-to-sequence generation framework.
The first sequence refers to the dynamic prompt, while the second sequence refers to the SEL sequence.
\subsubsection{Structured Extraction Language}
\begin{figure}[t]
    \centering
    \includegraphics[width=0.5\textwidth]{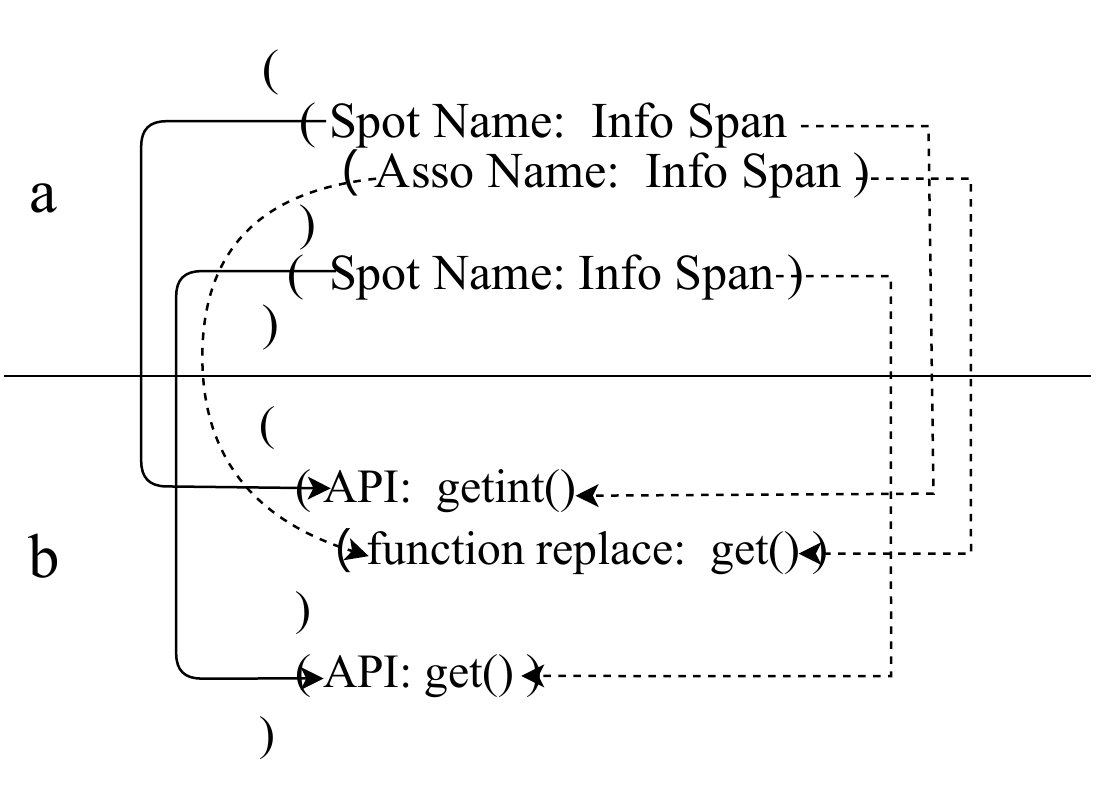}
    \caption{Specialization of Universal Structured Extraction Language.}
    \label{fig:sel_fig}
\end{figure}

SEL sequence is proposed to encode different information extraction structures via the hierarchical spotting-associating structure.
Fig.~\ref{fig:sel_fig}.a shows its universal format.
``Spot Name: Info Span'' denotes various entity type and the object of a specific entity type; ``Asso Name: Info Span'' denotes various relation types and the associated object of a specific relation type.
Fig.~\ref{fig:sel_fig}.b shows the concrete SEL sequence in our work.
``API: getint()'' represents that ``getint()'' is an API entity; ``function replace: get()'' represents that the relation between ``getint()'' and ``get()'' is ``function replace''.
From this concrete SEL sequence, we can extract API entities and relations  simultaneously as it unifies the structure of API entities and relations.
\subsubsection{SEL Sequence Generation}
We implement our API joint entity-relation extractor as the sequence-to-sequence generation framework:
\[\left[y_{1}, \ldots, y_{|y|}\right]=\mathrm{JE}(\left[p_{1}, \ldots, p_{|p|}\right])\tag{1}\]
where JE is a Transformer-based LLM, \(\left[p_{1}, \ldots, p_{|p|}\right]\) is the dynamic prompt, \(\left[y_{1}, \ldots, y_{|y|}\right]\) is the linearized SEL sequence that contains the API entities and relations to be extracted.
In this framework, we feed the dynamic prompt into the LLM (as shown in Fig.~\ref{fig:frame_fig}.II), and the LLM generates the SEL sequence (as shown in Fig.~\ref{fig:frame_fig}.III), from which we can obtain API entities and relations.
The dynamic prompt to the JE can also be written in the format described in Section~\ref{sec:procon}:
\[\begin{aligned}
\left[p_{1}, \ldots, p_{|p|}\right]
=& {[[\text { spot }], \ldots[\text { spot }] \ldots,} \\
& {[\text { asso }], \ldots,[\text { asso }] \ldots, } \\
& {\left.[\text { text }], x_{1}, x_{2}, \ldots, x_{|x|}\right] }
\end{aligned}\tag{2}\]
where
\(x=\left[x_{1}, \ldots, x_{|x|}\right]\) 
denotes the input text.

To better illustrate the framework's internal mechanics, an encoder-decoder-style architecture is introduced.
Given the dynamic prompt \(p\), JE computes the hidden representation
\(\mathbf{H}=\left[\mathbf{p}_{1}, \ldots, \mathbf{p}_{|p|}\right]\) of each token:
\[\mathbf{H}=\operatorname{Encoder}\left(p_{1}, \ldots, p_{|p|}\right)\tag{3}\]
where Encoder(·) is a Transformer encoder. Then JE decodes the prompt into a SEL sequence in an auto-regressive style.
At the step \(i\) of decoding, JE generates the \(i\)-th token \(y_{i}\) in the SEL sequence and the decoder state \(\mathbf{h}_{i}^{d}\) as following:
\[y_{i}, \mathbf{h}_{i}^{d}=\operatorname{Decoder}\left(\left[\mathbf{H}; \mathbf{h}_{1}^{d}, \ldots, \mathbf{h}_{i-1}^{d}\right]\right)\tag{4}\]
Decoder(·) is a Transformer decoder that predicts the conditional probability \(p\left(y_{i} \mid y<_{i},p\right)\) of token \(y_{i}\) until the end symbol $<$eos$>$ is output.

\subsection{Model Training}
This section describes data collection and augmentation, model training, which includes training a BERT-based classifier and fine-tuning a Transformer-based LLM.
\subsubsection{Data Collection}\label{sec:datacoll}
Given that the relation types we consider are all from a knowledge graph of Java APIs~\cite{9894095}, we randomly chose 5,000 Java-tagged posts from the Stack Overflow data dump~\footnote{\href{https://archive.org/download/stackexchange/}{Retrieved June
6, 2022 from https://archive.org/download/stackexchange/}}. 
Each post is accompanied by its answers and post tags (such as ``java'', ``arrays'', ``java.lang'').
We choose the most voted answers from the posts to ensure the quality of the training data, but we exclude code snippets and all HTML tags because the focus of our study is informal text.
All the answers are then splitted into sentences using spaCy~\footnote{\href{https://spacy.io}{https://spacy.io}}, yielding 28,140 sentences.
Every sentence is accompanied by multiple category tags from the post to which it belongs.
Then, for each sentence, we parse it into tokens using the software-specific tokenizer~\cite{Ye2016LearningTE} which preserves the integrity of an API mention.
\textit{iterator.remove()}, for example, is treated as a single token. 
Finally, we crawl all APIs in JDK 1.8~\footnote{\href{https://docs.oracle.com/javase/8/docs/api}{https://docs.oracle.com/javase/8/docs/api}}, and use these APIs to filter out the sentences containing API entities, as inspired by a previous study~\cite{Huo2022ARCLINAA}, based on the following criteria:
\begin{itemize} 
    \item 
    Because of the large number of morphological ambiguities, a token may be an API entity if it partially matches any of the crawled APIs (e.g., \textit{remove()} and \textit{java.util.Iterator.remove()}).
    \item
    Since API mentions usually end with ``\textit{()}'', the token is treated as an API entity if it contains ``\textit{()}''.
    \item
    API mentions typically include ``.'' to indicate a function call (e.g., \textit{iterator.remove()}, or \textit{l.remove()}); thus, if token contains ``.'', we consider it to be an API entity .
\end{itemize}

After filtering, we obtain 9,111 sentences that may contain API entities.
However, this is rough sentence filtering.
In order to do accurate sentence filtering, We invite 12 master students (all with more than five years Java experience) to examine the API entities and annotate the semantic relations between APIs in order to further verify whether these sentences contain API entities and the seven types of API relations we aim to extract.
We train the annotators prior to annotation to ensure that they can recognize these API relations in the text. 
After training, the annotators were divided into six groups, with two students from each group annotating the same content. 
After the annotation, we assign two authors to deal with the annotation results' conflicts, and the Cohen’s Kappa~\cite{Landis1977AnAO} coefficient is 0.859 (i.e., almost perfect agreement).
As a result, we get a total of 2917 sentences, with 2471 containing only entities and 446 containing both entities and relations.
\subsubsection{Data Augmentation}\label{sec:data aug}
To improve the AERJE’s ability to recognize API entities and relations from long and ambiguous sentences, we devise two data augmentation strategies to obtain more ambiguous sentences for model training.

\textbf{Morphology based Mutation.}
Inspired by~\cite{Huo2022ARCLINAA}, we change the form of each API entity in the sentence.
Specifically, we replace the API entity itself with the final piece of its fully qualified name. 
For example, \textit{iterator.remove()} is replaced with \textit{remove()} or \textit{remove}. 

\textbf{Verb based Mutation.}
We use spaCy to locate the verbs on which each API entity relies, and then replace those verbs with synonyms, as Liu et al.~\cite{Liu2022howto} do to obtain similar question titles.
As shown in the seventh sentence of Table~\ref{tab:table1_amb}, we replace ``read'' with ``load''.
However, because spaCy may not obtain the correct API entity, we must identify the dependency between the API entity's subtoken and the verb to ensure the mutation quality.
For example, there is a dependency between ``\textit{nextline}'' and ``read'', so we can reliably mutate ``read'' with synonyms.

Our data augmentation strategy does not include sentence pattern mutation~\cite{Liu2022howto}, which uses different sentence patterns to present the same API relation between the same API entities.
Unlike the morphology-based and verb-based mutation, this mutation is not reliable in software text which demands stricter semantics than general text. 
The sentence pattern mutation could result in sentence structure reconstruction, which would likely change the sentence semantics, contaminate the training data, and compromise AERJE training.
For example, the original sentence ``StringBuffer is synchronized, StringBuilder is not'' may be mutated into ``StringBuffer and StringBuilder differ in synchronized''.
The original sentence indicates that \textit{StringBuffer} is synchronized and \textit{StringBuilder} is asynchronous, but the mutated sentence does not specify who is synchronous or asynchronous.

We obtain 2,334 sentences as the initial training set and 583 sentences as the initial test set in an 8:2 ratio from the 2,917 sentences collected.
The number of sentences after applying the two data augmentation strategies to the initial training and test sets is 10,678 and 2,686, referred to as the final training set and the final test set, respectively.
This final training set is used to fine-tune the LLM-based extractor, and the final test set is used to test the fine-tuned extractor. 
Here, we split the sentences into training and testing sets and then mutated them. 
This ensures that the sentence before and after the mutation is in the same set, preventing the leaking of training data into the test set (e.g., one sentence in the training set and its mutation in the test set).
Furthermore, we obtain 1,639 sentences with both entities and relations as the classifier training set from the final training set.
Similarly, we obtain 387 sentences with both entities and relations as the classifier test set from the final test set.

\subsubsection{BERT-based Relation Classifier Training}
We choose BERT~\cite{Devlin2019BERTPO} as a relation classifier because its pre-training task (i.e., Next Sentence Prediction) is consistent with our task, both of which are classification tasks.
However, the implementation of relation classifier is not limited to BERT, we can also use TextCNN~\cite{Kim2014ConvolutionalNN} and FastText~\cite{Joulin2017BagOT}.
In our current implementation, we use the BERT-base classifier to classify each input sentence into N relation types.
Based on the N relation types, dynamic prompt generator generates the corresponding dynamic prompt.

A mask language model (BERT)~\cite{Devlin2019BERTPO} and a linear layer comprise the classifier.
Due to the seven API relation types, the linear layer's output dimension is set to 7.
We obtain the latent vector from the CLS token when we enter the sentence into BERT.
The latent vector obtained from the CLS token characterizes the sentence features better than other positions, resulting in better classification performance.
The latent vector is then fed into the linear layer, which produces a vector with seven dimensions, each corresponding to a relation type.
Finally, the classifier is trained on the classifier training set.
In back propagation, we use the cross-loss entropy to calculate the classifier's loss and adjust the BERT and linear layer parameters.
The loss function is formulated as follows, where
\(z=\left[z_{0}, \ldots, z_{C-1}\right]\)
represents the linear layer's output result, and C represents the sentence's label.
\[\operatorname{Loss}(z, c)=-z[c]+\log \left(\sum_{j=0}^{C-1} \exp (z[j])\right)\tag{5}\]

\subsubsection{LLM-based Extractor Fine-tuning}
We use the pre-trained T5-v1.1-large model~\cite{Raffel2020ExploringTL} as the LLM in our current implementation because T5's training objective aligns perfectly with our formulation of the API entity and relation extraction task as a sequence to sequence generation task.
Furthermore, studies~\cite{Jiang2021ExploringLE,Carmo2020PTT5PA} confirm that T5 is capable of capturing rich text information and demonstrate its effectiveness in a variety of downstream NLP tasks.
Our approach is not limited to T5, but can use any Transformer-based LLM.

In order to fine-tune T5, we convert each labeled sentence in the final training set into a SEL sequence (\textit{y}), then feed it into the dynamic prompt generator to obtain its dynamic prompt (\textit{p}), and finally construct the labeled corpus: \(\mathcal{D}_{\text {e }}= \{(p, y)\}\). 
On the labeled corpus, we fine-tune T5 for 50 epoch with batch size 10 using the Adam optimizer with a learning rate of 1e-4, linear scheduling with a warming up proportion of 6\%, and the teacher-forcing cross-entropy loss:
\[\mathcal{L}_{\mathrm{FT}}=\sum_{(p, y) \in \mathcal{D}_{\mathrm{e}}}-\log P\left(y \mid p ; \theta_{e}, \theta_{d}\right)\tag{6}\]
where \(\theta_{e}\) and \(\theta_{d}\) are the parameter of encoder and decoder, respectively.

\vspace{-2mm}
\section{EXPERIMENTAL SETUP}
This section starts with five questions about AERJE's performance, followed by a description of the experimental setup, which includes the dataset, baseline, and evaluation metrics.

\subsection{Rearch question}
\begin{itemize}
    \item 
    RQ1: Effectiveness of Data Augmentation
    \item
    RQ2: Optimal Num. of Relation Types for Dynamic Prompt
    \item
    RQ3: Joint Extraction Performance of AERJE
    \item
    RQ4: Generalization Ability of AERJE
    \item
    RQ5: AERJE’s Performance in Low-Resource Scenario
\end{itemize}

\subsection{Dataset}\label{sec:data_detail}
As described in section~\ref{sec:data aug}, there are three groups of data sets.
The first group refers to the sentences collected initially, some of which contain only entities and others contain both entities and relations.
\begin{itemize} 
    \item 
    The initial training set consists 2,334 sentences, of which 362 contain both entities and relations.
    \item
    The initial test set consists 583 sentences, of which 84 contain both entities and relations.
\end{itemize}

The second group refers to the sentences after applying the two data augmentation strategies, some of which contain only entities and others contain both entities and relations.
\begin{itemize}
    \item 
    The final training set with a total of 10,678 sentences, 1639 of which contain both entities and relations.
    \item
    The final test set with a total of 2,686 sentences,387 of which contain both entities and relations.
\end{itemize}

The third group refers to the sentences containing both entities and relations in the final training and testing sets.
\begin{itemize}
    \item 
    The classifier training set with a total of 1,639 sentences.
    \item
    The classifier test set with a total of 387 sentences.
\end{itemize}


\subsection{Baselines}\label{sec:variant}
Our AERJE is capable of API entity-relation joint extraction.
However, to the best of our knowledge, no previous work has focused on extracting both API entities and relations from unstructured texts at the same time. 
As a result, we can only compare AERJE with the existing work in the respective fields of API entity extraction and API relation extraction.

For API entity extraction, there are rule-based methods (such as regular Expressions~\cite{bacchelli2010linking,Treude2016AugmentingAD}), heuristic rule matching methods~\cite{9894095,liu2020generating,Ren2020APIMisuseDD}, and sequence-labeling based methods (such as ARCLIN ~\cite{Huo2022ARCLINAA} using BI-LSTM as encoder and CRF as decoder, APIReal~\cite{ye2018apireal} using only CRF).
Since the performance of the first two classes of methods is not as good as that of the last class of methods~\cite{Huo2022ARCLINAA}, we choose ARCLIN, APIReal as baselines.
We obtain them source code from Github~\footnote{\href{https://github.com/YintongHuo/ARCLIN}{https://github.com/YintongHuo/ARCLIN}}~\footnote{\href{https://github.com/baolingfeng/APIExing}{https://github.com/baolingfeng/APIExing}}, and label the API entities and non-entities in the sentences with the ``BIO'' tag (i.e., ``B'': the beginning of API entity segment, ``I'': the inside of API entity segment, ``O'': non-entity).
Then we create pairs of original sentences and labeled sentences to train these two baselines.
Finally, we the trained models on the final test set, from which we obtain API entities based on the ``BIO'' tag.

For API relation extraction, there are only rule matching methods that rely on API syntax~\cite{liu2020generating}, special-tag annotated relations~\cite{Li2018ImprovingAC}, or some ad-hoc relation phrases~\cite{9894095}.
It is very difficult to re-implement these methods due to the rule-design overhead.
Furthermore, it is impractical to apply these methods as we assume plain texts without any special annotations.
Instead, we implement a variant $AERJE_{w/oDPG}$ (${DPG}$ means the dynamic prompt generator), which uses a static prompt with all 7 relation types to evaluate the performance of the full-version of AERJE.

In addition, we also implement two other variants of AERJE as our baseline.
One is $AERJE_{single}$, which separate API entity and relation extraction into two independent tasks.
For API entity extraction, the prompt contains only ``[spot] API''.
For API relation extraction, the prompt contains only ``[asso] relation type''.
$AERJE_{single}$ still uses dynamic prompt in relation extraction.
After relation extraction, we merge the extracted entities and relations as the final results of $AERJE_{single}$. e.g., the extracted entities getint(), get() and relation function replace are  merged as (API: getint() (function replace: get())).
We compare $AERJE_{single}$ with AERJE to understand the effectiveness of joint entity-relation extraction.
Meanwhile, the entity extraction results of $AERJE_{single}$ is equivalent to fine-tuning pre-trained model for entity extraction, and its final results is equivalent to fine-tuning pre-trained model for relation extraction.
Therefore, $AERJE_{single}$ also reflects the capability of fine-tuning pre-trained model for entity and relation extraction separately.
Another variant is $AERJE_{base}$, which replace T5-v.1.1-large in AERJE with a smaller model backbone, i.e., T5-v1.1-base. 
We use it to explore the impact of large pre-trained language models on AERJE performance.
All variants use the same hyper-parameters as AERJE and remain constant across experimental scenarios.
Note that SEL used in AERJE has been demonstrated to be effective in the extraction task~\cite{Lu2022UnifiedSG}.
As such, we do not to verify the effectiveness of SEL in AERJE.

\subsection{Evaluation Metrics}
\vspace{-1mm}

We use Precision, Recall, and F1 score as metrics to evaluate the performance of AERJE and baseline models on our test set.
Precision means what percentage of API entities and relations extracted are correct, recall means what percentage of the real API entities and relations are extracted, and F1 score is the harmonic mean of precision and recall.
It is important to note that the relation is only correct if the relation type and corresponding entities are both correct. 
In context of our work, we are not concerned with the top-N relation classification accuracy.
As long as the top-N includes relevant relation types, the extractor does not care about the order of these relation types.
Furthermore, a sentence may have 2 or more relations, which renders the top-1 accuracy irrelevant.
Finally, as the extractor has the capability of ruling out irrelevant relation types in the prompt, it is also not necessary to evaluate the classification precision and recall at N.

\section{EXPERIMENTAL RESULTS}
This section delves into five research questions to evaluate and discuss the AERJE's performance.

\subsection{RQ1: Effectiveness of Data Augmentation}
\subsubsection{Motivation}
To reduce manual labeling effort and improve model training, we devise two data augmentation strategies.
We want to investigate if ambiguous but correctly annotated sentences obtained through two data augmentation strategies could improve AERJE's discriminative capability for extracting API entities and relations, in order to demonstrate the effectiveness of two data augmentation strategies.

\subsubsection{Methodology}
We set up two scenarios $AERJE_{w/oDA}$ and $AERJE_{wDA}$ (${DA}$ means the data augmentation).
$AERJE_{w/oDA}$ is trained on the initial training set, while $AERJE_{wDA}$ is trained on the final training set.
Both $AERJE_{w/oDA}$ and $AERJE_{wDA}$are tested on the same final test set.
This setting allows us to compare the effectiveness of data augmentation.

\subsubsection{Result}
Table~\ref{tab:rq1_result} shows the experimental results.
In terms of API entity extraction, $AERJE_{wDA}$  has precision, recall, and F1-scores of 97.57\%, 95.48\%, and 96.51\%, while $AERJE_{w/oDA}$ has precision, recall, and F1-scores of 95.11\%, 92.19\%, and 93.63\%.
$AERJE_{wDA}$'s precision, recall, and F1-score are all higher than $AERJE_{w/oDA}$'s, with $AERJE_{wDA}$'s recall and F1-score being about 3\% higher.

In terms of API relation extraction, $AERJE_{wDA}$ has precision, recall, and F1-score of 86.54\%, 76.48\%, and 81.20\%, while $AERJE_{w/oDA}$ has precision, recall, and F1-score of 77.71\%, 75.66\%, and 76.67\%.
$AERJE_{wDA}$'s precision, recall, and F1-score are all higher than $AERJE_{w/oDA}$'s.
The precision of $AERJE_{wDA}$ is 8.83\% higher than that of $AERJE_{w/oDA}$, and the F1-score of $AERJE_{wDA}$ is 4.53\% higher than that of $AERJE_{w/oDA}$.
This demonstrates that fine-tuning $AERJE_{wDA}$ using a large number of ambiguous sentences with API relations benefits $AERJE_{wDA}$ to distinguish between relations and non-relations, as well as between correct and incorrect relations.
In contrast, $AERJE_{w/oDA}$ has not been fine-tuned on ambiguous sentences and thus does not perform as well as $AERJE_{wDA}$.
For example, an ambiguous sentence ``you want to read up on processbuilder to launch the exe file and then waitfor() to wait until the process is complete''.
$AERJE_{wDA}$ correctly extracts two API entities, \textit{ProcBuilder} and \textit{waitfor()}, as well as the ``logic constraint'' relation between them, from the sentence. 
In contrast, $AERJE_{w/oDA}$ only extracts one API \textit{waitfor()} from the sentence.
This shows AERJE's capability to extract API entities and relations from ambiguous sentences can be improved by fine-tuning with the augmentated data.

\vspace{2mm}
\noindent\fbox{
\begin{minipage}{13.5cm}{
AERJE's discriminative capability for API entities and relations can be improved by fine-tuning it with ambiguous but correctly labeled sentences obtained through the data augmentation strategies.} \end{minipage}}

\begin{table}[t]
\caption{Impact of data augmentation strategy on AERJE}
\label{tab:rq1_result}
\begin{threeparttable}
\begin{tabular}{c|ccc|ccc}
\hline
\multirow{2}{*}{Strategy} & \multicolumn{3}{c|}{Entity} & \multicolumn{3}{c}{Relation} \\ \cline{2-7} 
                          & P       & R       & F1      & P        & R       & F1      \\ \hline
$AERJE_{wDA}$             & \textbf{97.57}   & \textbf{95.48}   & \textbf{96.51}   & \textbf{86.54}    & \textbf{76.48}   & \textbf{81.20}   \\ \hline
$AERJE_{w/oDA}$           & 95.11   & 92.19   & 93.63   & 77.71    & 75.66   & 76.67   \\ \hline
\end{tabular}
\end{threeparttable}
\end{table}

\subsection{RQ2: Optimal Num. of Relation Types for Dynamic Prompt}
\subsubsection{Motivation}
As described in section~\ref{sec:dypro}, given an input sentence, the dynamic prompt generator employs the BERT-based classifier to predict a set of candidate relation types, which are then included in the dynamic prompt to guide the subsequent joint entity-relation extractor.
In this RQ, we would like to investigate how many candidate relation types (i.e., top-N classifier results) can provide the most effective guidance to the extractor.

\subsubsection{Methodology}
We exhaust all cases of N values (from 1 to 6) in the dynamic prompt generator, then fine-tune AERJE on the same final training set and test it on the same final test set to select the most appropriate N value based on experimental results.
We do not test N=7 because it is essentially the static prompt with all seven relation types (i.e., $AERJE_{w/oDPG}$ studied in RQ3).

\subsubsection{Result}
As shown in Table~\ref{tab:rq2}, changing the N value has small effect on entity extraction because N represents the number of relation types in the dynamic prompt which does not directly affect entity extraction.
At N=3, AERJE achieves the marginally best F1-score 96.51\% for API entity extraction.

For relation extraction, changing the N value has larger effect on both precision and recall.
As N increases, both precision and recall improve until N=3.
When N=3, the precision, recall and F1-score of AERJE reaches the highest 86.54\%, 76.48\% and 81.20\%, respectively.
This means that the correct API relation type is most likely covered in the top-3 candidate relations predicted by the classifier.
When N is less than 3, however, the F1-score of AERJE in relation extraction decreases because the top-N candidate relations may miss the correct relation type.
Here is an example: ``A TreeMap has the same limitation (as does a HashMap, which also breaks when the hashcode of its elements changes after insertion)''. 
When N=2, classifier predicts two relations between \textit{TreeMap} and \textit{Hashmap}, including ``behavior difference'' and ``logic constraint''
, but ignores the ``function similarity'' relation.
This ignored relation is at the third relation predicted by the classifier.
However, when N is greater than 3, the F1-score of AERJE in relation extraction decreases because the dynamic prompt may contain some incorrect relation types, which may mislead the extractor.
This misleading effect has bigger impact on precision than on recall.

\vspace{2mm}
\noindent\fbox{
\begin{minipage}{13.5cm}{
The optimal number of relation types for dynamic prompt should be set to 3. This not only ensures that the majority of the correct relation types appear in the dynamic prompts, but it also prevents the dynamic prompts from containing too many noise relation types which may make the model sacrifice precision for recall.} \end{minipage}}

\begin{table}[t]
\caption{Model results for different values of N}
\label{tab:rq2}
 \begin{tabular}{c|ccc|ccc}
\hline
\multirow{2}{*}{top-N} & \multicolumn{3}{c|}{Entity} & \multicolumn{3}{c}{Relation} \\ \cline{2-7} 
                       & P       & R       & F1      & P        & R       & F1      \\ \hline
1                      & 96.88   & 94.23   & 95.54   & 75.92    & 71.92   & 73.87   \\ \hline
2                      & 97.04   & 95.18   & 96.10   & 77.75    & 73.80   & 75.72   \\ \hline
3                      & 97.57 & \textbf{95.48}   & \textbf{96.51}   & \textbf{86.54}   & \textbf{76.48}   & \textbf{81.20}   \\ \hline
4                      & \textbf{97.84}   & 94.39   & 96.08   & 83.51    & 73.22   & 78.03   \\ \hline
5                      & 96.72   & 94.39   & 95.54   & 77.90    & 73.30   & 75.53   \\ \hline
6                      & 96.44   & 94.75   & 95.59   & 75.35    & 72.61  & 73.95   \\ \hline
\end{tabular}
\end{table}

\subsection{RQ3: Joint Extraction Performance of AERJE}
\subsubsection{Motivation}
We would like to evaluate AERJE's performance in API entity and relation joint extraction, compared with the state-of-the-art methods for API entity extraction and API relation extraction.
Note that only our AERJE can achieve joint API entity and relation extraction.

\subsubsection{Methodology}
AERJE is compared to APIReal and ARCLIN for API entity extraction, and three variant models (i.e., $AERJE_{w/oDPG}$, $AERJE_{single}$, and $AERJE_{base}$) for both API entity and relation extraction.
Note that the entity and relation extraction results by $AERJE_{single}$ represents the capability of fine-tuning the pre-trained model for the two tasks separately.
All models are trained and tested on the same final training and test sets.
Details on configuration can be found in Section~\ref{sec:variant}.

\begin{table}[t]\small
\caption{Comparison of Overall Performance}
\vspace{-2mm}
\label{tab:rq3}
\begin{threeparttable}
\begin{tabular}{c|ccc|ccc}
\hline
\multirow{2}{*}{Model} & \multicolumn{3}{c|}{Entity} & \multicolumn{3}{c}{Relation} \\ \cline{2-7} 
                       & P       & R       & F1      & P        & R       & F1      \\ \hline
APIReal                & 89.13   & 88.90   & 89.01   & -        & -       & -       \\ \hline
ARCLIN                 & 94.76   & 87.17   & 90.81   & -        & -       & -       \\ \hline
AERJE                  & 97.57   & 95.48   & 96.51   & \textbf{86.54}    & \textbf{76.48}   & \textbf{81.20}   \\ \hline
$AERJE_{single}$       & \textbf{98.03}  & 95.38   & \textbf{96.69}   & 82.83    & 67.47   & 74.37   \\ \hline
$AERJE_{w/oDPG}$       & 97.52   & \textbf{95.78}   & 96.64   & 75.38    & 70.62   & 72.92   \\ \hline
$AERJE_{base}$         & 96.39   & 95.15   & 95.77   & 75.97    & 70.28   & 73.01   \\ \hline
\end{tabular}
\end{threeparttable}
\vspace{-6mm}
\end{table}

\subsubsection{Result}
Table~\ref{tab:rq3} shows the evaluation result of AERJE and five baselines on final test sets.
We see that AERJE's F1-score is higher 7.5\% than APIReal's F1-score and 5.7\% than ARCLIN's F1-score on API entity extraction.
Compared with the three variant models, AERJE's F1-score for API entity extraction is only slightly lower (0.18\%) than the best performer (i.e., 96.69\% by $AERJE_{single}$), but AERJE's F1-score for relation extraction is 6.83\% higher than that of the second best performer (74.37\% by $AERJE_{single}$).

For APIReal and ARCLIN performance on API entity extraction, both AERJE and it variant models outperform them largely.
This superior performance is due to the backbone large pre-trained language models (T5) in AERJE.
During the pre-training, T5 learns linguistic and semantic knowledge in text and has powerful abilities in word and sentence representations.
Through fine-tuning, the semantic knowledge packed in the T5 can be transferred to the downstream tasks and benefit API entity extraction.

The amount of knowledge in the T5 also affects the AERJE's performance on API entity and relation extraction.
Compared with AERJE, the F1-score of $AERJE_{base}$ is reduced by 0.74\% and 8.19\% in API entity extraction and API relation extraction, respectively.
The decrease of $AERJE_{base}$'s F1 score on API entity extraction is very small compared with the decrease on API relation extraction.
It is because the number of sentences containing API entities in the final training set is 6 times more than the number of sentences containing both API entity and relation (i.e., 10,678 vs 1,639).
Sufficient fine-tuning data for API entity extraction allows the basic T5 model to achieve the equivalent performance on API entity extraction as the large T5.
In contrast, the relation extraction is more complex than the entity extraction, and the amount of fine-tuning data is smaller.
In such case, the basic T5 cannot compete with the large T5.

For $AERJE_{single}$ and AERJE, they achieve almost the same entity extraction performance. 
However, in terms of API relation extraction, AERJE’s F1-score (81.20\%), precision (86.54\%) and recall (76.48\%) are much higher than $AERJE_{single}$’s F1-score (74.37\%), precision (82.83\%) and recall (67.47\%), respectively.
This suggests that fine-tuning pre-trained model for API entity extraction individually or jointly with API relation extraction does not affect the quality of API entity extraction.
But joint entity and relation extraction is much more effective for the relation extraction task than fine-tuning the model just for the relation extraction.

For $AERJE_{w/oDPG}$ and AERJE, they also achieve almost the same entity extraction performance.
This is due to the fact that dynamic prompt only affects the relation type, not the entity type.
In terms of API relation extraction, AERJE's precision (86.54\%), recall (76.48\%) and F1-score (81.20\%) are higher than $AERJE_{w/oDPG}$'s precision (75.38\%), recall (70.62\%) and F1-score (72.92\%), respectively.
This is because $AERJE_{w/oDPG}$ uses the same static prompt that includes all seven relation types for all input sentences.
The more types of relations there are in the prompt, the more noise the prompt is, and the more difficult it is for AERJE to identify and extract the correct relations in the input sentence. 
In contrast, AERJE's use of dynamic prompt reduces the number of relation types to recognize, improving its ability to extract API relations.

\vspace{1mm}
\noindent\fbox{
\begin{minipage}{13.5cm}{
Standing on the shoulder of large pre-trained language model (T5), AERJE outperforms traditional sequence labeling models for API entity extraction.
Dynamic prompt has no impact on API entity extraction, but can largely boost the performance of API relation extraction.
Fine-tuning the pre-trained model jointly is much more effective than fine-tuning the model just for one task, which makes joint entity-relation extraction more accurate on both tasks than separate entity and relation extraction.} \end{minipage}}

\subsection{RQ4: Generalization Ability of AERJE}
\subsubsection{Motivation}
Each API comes with its own API package, which often have different forms.
Furthermore, as APIs from different packages support diverse functionalities, the texts in which they appear may be different in content and linguistic properties.
It is impossible for AERJE to see all API packages during fine-tuning. 
In this RQ, we want to investigate if AERJE can recognize APIs and their relations from the API packages that it does not see during fine-tuning.

\subsubsection{Methodology}
In order to collect as much data from different packages as possible, we combine the final training set and the final test set into a new data set with a total of 13,364 sentences.
Every sentence, as stated in Section~\ref{sec:datacoll}, is accompanied by multiple post tags, some of which show the relationship between the sentence and the API package.
For example, the tag ``io'' is associated with the package name ``java.io''.
Therefore, we filter out sentences with package names by matching each tag of a sentence to any JDK 1.8 package name.
Here is a partial match, which means it matches a portion of the package name, for example, ``swing'' can match ``javax.swing''.
And then we pool the package names that appear with the sentences and select the three package names that appear the most frequently (i.e., javax.swing, java.io, and java.util).
Finally, we gather 1651 sentences whose tags match these three package names.

To ensure the correctness of the sentences obtained through approximate match, we invite six students (who have previously participated in annotation) and divide them into three groups to annotate sentences from three different packages.
Two students in each group annotate the same sentences. 
They independently determine whether the API entities in each sentence are from the specific package (i.e., java.io, java.util, javax.swing).
Here is an example ``you can use lines() method in BufferedRead'' for java.io package. 
The sentence is annotated as True, since the API entities \textit{line()} and \textit{BufferedRead} only correspond to java.io.
Instead, if any API entity in the sentence do not belong to specific package, the sentence is annotated as False.
Then we assign an author to handle conflicts between the group members.
Finally, we obtain 999 sentences that strictly matched these packages names.
Cohen’s Kappa~\cite{Landis1977AnAO} coefficient is 0.795 (i.e., substantial agreement).
The data details for each package are as follows:
\begin{itemize} 
    \item 
    The java.io dataset has 235 sentences, 51 of which contain both entities and relations.
    12 of the 51 sentences are non-augmented sentences.
    \item
    The javax.swing dataset has 435 sentences, 76 of which contain both entities and relations. 
    14 of the 76 sentences are non-augmented sentences. 
    \item
    The java.util dataset has 329 sentences, 68 of which contain both entities and relations.
    18 of the 68 sentences are non-augmented sentences.
\end{itemize}


Our AERJE and baseline models are all trained on one of the three datasets and tested on the two others.
As AERJE outperforms its variants. We don't consider these variants here.

\subsubsection{Result}
Table~\ref{tab:rq4} shows the results that reflect each model's generalization ability.
For API entity extraction, AERJE's F1-score achieves 95.05\%, when trained on the java.util dataset, far exceeding APIReal's F1-score (61.27\%) and ARCLIN's F1-score (58.50\%). 
We attribute this to the underlying LLM on which AERJE is built.
As Qiu et al.~\cite{Qiu2020PretrainedMF} show, LLM provides better model initialization, which usually leads to better generalization performance on the target tasks.
Similar observations can be made for training the models on the java.io dataset and the javax.swing dataset.
Generally, ARCLIN and APIReal may perform well on either precision or recall, but not both and thus poor F1-score.
In contrast, AERJE is very stable with much better precision and recall and with only small fluctuations in F1-scores across the experiments.

For API relation extraction, AERJE's F1-score is 40.98\%, 79.99\% and 68.48\%  when trained on the java.io, java.util and javax.swing datasets, respectively.
In the across-package training-testing setting, the performance of AERJE degrades, compared with the non-across-package setting (see Table~\ref{tab:rq3}).
However, when trained on the java.util dataset, AERJE's F1-score (79.99\%) is only about 1\% less than non-across-package setting (81.20\%).
This suggests that AERJE is capable of dealing with the data drift across different packages.
In addition, different across-package training-testing settings also bring different results.
When using java.util for training AERJE, its F1-score is about 39\% higher than the F1-score of AERJE trained on java.io. 
First, due to java.io having fewer sentences with relations than java.util (51 vs 68).
Second, java.io data has fewer non-augmented sentences with relations than java.util (12 vs 18), which makes java.io data less diverse than java.util.

\vspace{2mm}
\noindent\fbox{
\begin{minipage}{13.5cm}{Our AERJE has a strong generalization ability in face of the data drift across different API packages. This ability comes from the generalization ability of the underlying LLM.} 
\end{minipage}}
\subsection{RQ5: AERJE's Performance in Low-Resource Scenario}
\subsubsection{Motivation}
Labor overhead means that the data available for training is limited.
In this RQ, we want to investigate how well AERJE perform when trained with the extremely small amount of training data.
\subsubsection{Methodology}
We conduct a K-shot experiment, where K can be 1, 5, or 10.
To begin the K-shot experiment, we randomly select K sentences from the final training set for each relation type.
Then we choose K sentences at random from the final training set that contain only entities but no relations.
This yields a training set containing 8*k sentences.
Finally, we train our AERJE and baseline models on this training set and test them on the final test set. 
Note that, to avoid the influence of random sampling, we repeat each K-shot experiment ten times with different samples.

\begin{table*}[t]
\caption{Comparison of Generalization Ability}
\label{tab:rq4}
\resizebox{\linewidth}{!}{
    \begin{tabular}{c|cccccc|cccccc|cccccc}
    \hline
    \multirow{3}{*}{Model} & \multicolumn{6}{c|}{java.io}                                               & \multicolumn{6}{c|}{java.util}                                             & \multicolumn{6}{c}{javax.swing}                                           \\ \cline{2-19} 
                          & \multicolumn{3}{c|}{Entity}                & \multicolumn{3}{c|}{Relation} & \multicolumn{3}{c|}{Entity}                & \multicolumn{3}{c|}{Relation} & \multicolumn{3}{c|}{Entity}                & \multicolumn{3}{c}{Relation} \\ \cline{2-19} 
                          & P     & R     & \multicolumn{1}{c|}{F1}    & P         & R       & F1      & P     & R     & \multicolumn{1}{c|}{F1}    & P         & R         & F1    & P     & R     & \multicolumn{1}{c|}{F1}    & P        & R       & F1      \\ \hline
    APIReal                & 85.02 & 36.97 & \multicolumn{1}{c|}{51.53} & -         & -       & -       & 98.11 & 44.54 & \multicolumn{1}{c|}{61.27} & -         & -         & -     & \textbf{98.99} & 25.62 & \multicolumn{1}{c|}{40.70}  & -        & -       & -       \\ \hline
    ARCLIN                 & \textbf{95.93} & 70.84 & \multicolumn{1}{c|}{81.50}  & -         & -       & -       & \textbf{98.55} & 41.60  & \multicolumn{1}{c|}{58.50}  & -         & -         & -     & 98.64 & 56.57 & \multicolumn{1}{c|}{71.90}  & -        & -       & -       \\ \hline
    AERJE                  & 92.00    & \textbf{89.35} & \multicolumn{1}{c|}{\textbf{90.66}} & 45.87    & 37.03   & 40.98   & 95.17 &\textbf{94.93} & \multicolumn{1}{c|}{\textbf{95.05}} & 78.68     & 81.35     & 79.99    & 93.89 & \textbf{89.91} & \multicolumn{1}{c|}{\textbf{91.86}} & 96.92    & 52.94   & 68.48   \\ \hline
    \end{tabular}}
\end{table*}

\begin{table*}[t]
\caption{Experimental results in a low-resource scenario}
\label{tab:rq5}
\resizebox{\linewidth}{!}{
\begin{tabular}{c|cccccc|cccccc|cccccc}
\hline
\multirow{3}{*}{Model} & \multicolumn{6}{c|}{1-Shot}                                                & \multicolumn{6}{c|}{5-Shot}                                                 & \multicolumn{6}{c}{10-Shot}                                               \\ \cline{2-19} 
                      & \multicolumn{3}{c|}{Entity}                & \multicolumn{3}{c|}{Relation} & \multicolumn{3}{c|}{Entity}                 & \multicolumn{3}{c|}{Relation} & \multicolumn{3}{c|}{Entity}                & \multicolumn{3}{c}{Relation} \\ \cline{2-19} 
                      & P     & R     & \multicolumn{1}{c|}{F1}    & P       & R        & F1       & P      & R     & \multicolumn{1}{c|}{F1}    & P        & R        & F1      & P     & R     & \multicolumn{1}{c|}{F1}    & P      & R        & F1       \\ \hline
APIReal                & \textbf{80.30}  & 15.92 & \multicolumn{1}{c|}{26.57} & -       & -        & -        & \textbf{86.17}  & 60.67 & \multicolumn{1}{c|}{71.20}  & -        & -        & -       & \textbf{83.94} & 68.59 & \multicolumn{1}{c|}{75.49} & -      & -        & -        \\ \hline
ARCLIN                 & 55.07 & 62.38 & \multicolumn{1}{c|}{58.50} & -       & -        & -        & 74.64 & 72.91 & \multicolumn{1}{c|}{73.76}  & -        & -        & -       & 83.58 & 75.52 & \multicolumn{1}{c|}{79.34} & -      & -        & -        \\ \hline
AERJE                  & 72.76 & \textbf{85.06} & \multicolumn{1}{c|}{\textbf{78.43}} & 9.34    & 44.49    & 15.44    & 79.09  & \textbf{91.62} & \multicolumn{1}{c|}{\textbf{84.90}} & 31.68    & 65.96    & 42.80   & 82.47 & \textbf{93.74} & \multicolumn{1}{c|}{\textbf{87.74}} & 35.00     & 72.94    & 47.30    \\ \hline
\end{tabular}}
\end{table*}

\subsubsection{Result}
For API entity extraction, Table~\ref{tab:rq5} shows the performance of each model in three low-resource scenarios (i.e., 1-shot, 5-shot, and 10-shot) where AERJE significantly outperforms APIReal and ARCLIN.
Especially, in the 1-shot scenario, AERJE's F1-score is 78.43\%, which is significantly higher than APIReal's (26.57\%) and ARCLIN's (58.50\%). 
Compared to APIReal and ARCLIN, the LLM-based AERJE has a large amount of prior knowledge from the LLM pre-training.
As the fine-tuning shot increases, the accuracy of AERJE improves fast, especially on F1-score, reaching the F1-score 84.90\% at 5-shot and 87.74\% at 10-shot.

For API relation extraction, in the 1-shot scenario, AERJE does not perform well, but it still magically achieves the recall 44.49\%.
However, with only 4 more examples (at 5-shot), the F1-score of AERJE significantly increases from below 16\% at 1-shot to about 43\% at 5-shot.
This suggests that the underlying LLM can quickly adapt to the API relation extraction task that it does not see during pre-training with only a few examples.
In the few-shot setting, we see that precision is much more difficult to improve than recall.
It could be due to the ambiguities of relations, i.e. the same type of relation can be expressed in very different forms (as shown in table~\ref{tab:table1_amb}), while different types of relations may be expressed in the similar forms (e.g., ``API1 be ADJ to API2'' represents function similarity or function opposite relation).
With only a few examples of each type of relation, it makes learning to distinguish between them more difficult.
Furthermore, AERJE's F1-score for entity extraction is 78.43\% at one-shot, while its F1-score for relation extraction is only 15.44\%.
The primary cause for this is that the training set of one-shot contains almost all API entity ambiguity types but only a few API relation ambiguity types.
As a result, relation extraction is more difficult than entity extraction.

\vspace{2mm}
\noindent\fbox{\begin{minipage}{13.5cm}{
AERJE can quickly adapt the underlying LLM to the API entity and relation extraction tasks with only a small number of fine-tuning data. Prior knowledge in LLM enables this quick adaptation. 
Relation extraction is much harder than entity extraction in the few-shot setting.
} \end{minipage}}
\section{DISCUSSION}

The major threat to internal validity is the manual labeling of training and testing datasets.
Incorrect human labels could harm modeling training and testing.
To mitigate this threat, we invited two students to annotate the same content and assigned an author to resolve disagreements in the labeling results.
However, even humans can't always tell if a token references an API, especially when it comes to common nouns that reference basic computing concepts, such as \textit{policy} and \textit{time}, which can be either basic noun concepts or APIs (\textit{java.security.policy} class, \textit{java.time} package).
We take a conservative strategy, i.e., common nouns as API entities, unless both annotators agree.

The threat to external validity is three-fold.
The first external threat is that we only collect data on Stack Overflow.
Although our model performed well on the SO data set, we intend further to validate its generalization performance in the other data sources (e.g., Java Tutorial\footnote{\href{http://www.java2s.com}{http://www.java2s.com}}, SitePoint\footnote{\href{https://www.sitepoint.com/}{https://www.sitepoint.com/}}, and Reddit\footnote{\href{https://www.reddit.com}{https://www.reddit.com}}).
The second external threat is that AERJE has only been tested on Java packages. 
We chose Java because previous work~\cite{liu2020generating,Ren2020APIMisuseDD,9894095} has demonstrated how difficult it is to extract these API entities and relations from it. 
In the future, we plan to expand AERJE to other programming languages (such as Python and C\#).
The third external threat stems from two AERJE components: the BERT-based classifier and the T5-based extractor.
There are numerous alternative models for both components of the model.
TextCNN~\cite{Kim2014ConvolutionalNN} and FastText~\cite{Joulin2017BagOT} can be used to build the classifier. 
It is possible to use BART~\cite{Lewis2020BARTDS} and GPT-3~\cite{Brown2020LanguageMA} to implement the extractor.
In the future, we will compare two AERJE components with alternative models to determine the best performing model.
\section{related work}
API entity and relation extraction is a fundamental work in software engineering.
It is useful in the construction of knowledge graphs;
extracted structured API knowledge can help with many software engineering tasks such as API linking~\cite{ye2018apireal,yin2021api,Dagenais2012RecoveringTL,Treude2016AugmentingAD}, API misuse detection~\cite{Ren2020APIMisuseDD}, API recommendation~\cite{huang2018api,Xie2020APIMR}, and API comparison~\cite{liu2020generating}.
This section describes the methods for extracting API entities and relations from unstructured text.

Bacchelli~\cite{bacchelli2010linking,Bacchelli2009BenchmarkingLT} and Dagenais~\cite{Dagenais2012RecoveringTL} detect class and method mentions in developer emails, documentation and forum posts using regular expressions of distinct orthographic features.
Ren~\cite{Ren2020APIMisuseDD}, Huang~\cite{9894095}, and Liu~\cite{liu2020generating} extract entities from SO posts using the HTML $<$code$>$ tag. 
Bacchelli et al.~\cite{Bacchelli2011ExtractingSD} extract coarse-grained structured code fragments from natural language text with island parsing.
Huang~\cite{9894095}, and Liu~\cite{liu2020generating} extract semantic relations between entities based on syntactic patterns.
However, their API entity and relation extraction method from natural language text relies on unique orthographic features of APIs, and suffer from the rule design overhead.

To mitigate the overhead of rule design, researchers extract API entities using machine learning methods.
Ye et al.~\cite{ye2018apireal} propose APIReal, which uses CRF to identify API entities.
They label the API entities and non-entities in the sentence with the ``BIO'' tag and form the pair of the labeled sequence and the sequence.
They then train CRF on these pairs, and use the trained CRF to label the input text, from which they obtain API entities with the ``BI'' or ``B'' tag.
Huo et al.~\cite{Huo2022ARCLINAA}, on the other hand, propose ARCLIN, which uses BI-LSTM as encoder and CRF as decoder to identify API entities, rather than just CRF.
However, these methods suffer from data labeling overhead because preparing a large number of high-quality training data for these sequence labeling models is unrealistic.

To solve the two overhead issues mentioned above, researchers use LLM to extract entities.
Li et al.~\cite{Li2020AUM} use BERT and Yan et al.~\cite{Yan2021NamedER} use XLNet~\cite{Yang2019XLNetGA} to extract entities in the natural language domain.
These models, however, are limited to a single natural language processing task, i.e., the entity extraction only.
In order to realize joint extraction of multiple tasks, researchers propose LLM-based unified architectural models, such as UIE~\cite{Lu2022UnifiedSG} and OpenUE~\cite{Zhang2020OpenUEAO}.
In particular, UIE proposes SEL to encode different information extraction structures via the hierarchical spotting-associating structure.
Motivated by this, we consider adapting UIE to the joint API entity-relation extraction.
However, UIE is not good at dealing with complex sentences, particularly long and ambiguous sentences containing API entities and various relations, because UIE has only one static prompt to identify all types of API relations.
As a result, when confronted with ambiguous sentences, the more relation types to recognize, the more noise interference, and the lower the UIE recognition rate.
In contrast, we propose LLM-based AERJE, which extracts API entities and relations from unstructured complex sentences at the same time.
Different from UIE, our dynamic prompt design could generate a small number of potentially relevant relations for input text to eliminate noise interference and lessens the difficulty of API relation extraction.
\section{conclusion and future work}
In this paper, we are the first to formulate heterogeneous API extraction and API relation extraction task as a sequence-to-sequence task, and proposes AERJE to extract API entities and relations from unstructured text simultaneously using pre-trained LLM and dynamic prompt learning.
The systematic evaluation of AERJE is conducted on a set of long and ambiguous sentences from Stack Overflow.
The experimental results show that AERJE's ability to extract API entities and relations can be activated with a small amount of data, allowing it to accurately identify API entities and relations from complex text that the model has never seen during fine-tuning.
In the future, we will carry out the plans mentioned in the discussion and apply AERJE to any software engineering task supported by API entity and relation extraction, such as API linking, API search, and API recommendation.
\begin{acks}
The work is partly supported by the National Nature Science Foundation of China under Grant (Nos.62262031, 61902162), the Nature Science Foundation of Jiangxi Province (20202BAB202015), the Central Guided Local Science and Technology Development Special Project (20222ZDH04090), the Graduate Innovative Special Fund Projects of Jiangxi Province (YC2021-S308, YC2022-S258).
\end{acks}

\normalem
\bibliography{sample-base}

\begin{thebibliography}{10}

\bibitem{9894095}
Qing Huang, Zhiqiang Yuan, Zhenchang Xing, Zhengkang Zuo, Changjing Wang, and
  Xin Xia.
\newblock 1+1$>$2: Programming know-what and know-how knowledge fusion,
  semantic enrichment and coherent application.
\newblock {\em IEEE Transactions on Services Computing}, pages 1--14, 2022.

\bibitem{ye2018apireal}
Deheng Ye, Lingfeng Bao, Zhenchang Xing, and Shang-Wei Lin.
\newblock Apireal: an api recognition and linking approach for online developer
  forums.
\newblock {\em Empirical Software Engineering}, 23(6):3129--3160, 2018.

\bibitem{Dagenais2012RecoveringTL}
Barth{\'e}l{\'e}my Dagenais and Martin~P. Robillard.
\newblock Recovering traceability links between an api and its learning
  resources.
\newblock {\em 2012 34th International Conference on Software Engineering
  (ICSE)}, pages 47--57, 2012.

\bibitem{huang2018api}
Qiao Huang, Xin Xia, Zhenchang Xing, David Lo, and Xinyu Wang.
\newblock Api method recommendation without worrying about the task-api
  knowledge gap.
\newblock In {\em 2018 33rd IEEE/ACM International Conference on Automated
  Software Engineering (ASE)}, pages 293--304. IEEE, 2018.

\bibitem{Xie2020APIMR}
Wenkai Xie, Xin Peng, Mingwei Liu, Christoph Treude, Zhenchang Xing, Xiaoxin
  Zhang, and Wenyun Zhao.
\newblock Api method recommendation via explicit matching of functionality verb
  phrases.
\newblock {\em Proceedings of the 28th ACM Joint Meeting on European Software
  Engineering Conference and Symposium on the Foundations of Software
  Engineering}, 2020.

\bibitem{liu2020generating}
Yang Liu, Mingwei Liu, Xin Peng, Christoph Treude, Zhenchang Xing, and Xiaoxin
  Zhang.
\newblock Generating concept based api element comparison using a knowledge
  graph.
\newblock In {\em Proceedings of the 35th IEEE/ACM International Conference on
  Automated Software Engineering}, pages 834--845, 2020.

\bibitem{bacchelli2010linking}
Alberto Bacchelli, Michele Lanza, and Romain Robbes.
\newblock Linking e-mails and source code artifacts.
\newblock In {\em Proceedings of the 32nd ACM/IEEE International Conference on
  Software Engineering-Volume 1}, pages 375--384, 2010.

\bibitem{Treude2016AugmentingAD}
Christoph Treude and Martin~P. Robillard.
\newblock Augmenting api documentation with insights from stack overflow.
\newblock {\em 2016 IEEE/ACM 38th International Conference on Software
  Engineering (ICSE)}, pages 392--403, 2016.

\bibitem{Rigby2013DiscoveringEC}
Peter~C. Rigby and Martin~P. Robillard.
\newblock Discovering essential code elements in informal documentation.
\newblock {\em 2013 35th International Conference on Software Engineering
  (ICSE)}, pages 832--841, 2013.

\bibitem{Bacchelli2011ExtractingSD}
Alberto Bacchelli, Anthony Cleve, Michele Lanza, and Andrea Mocci.
\newblock Extracting structured data from natural language documents with
  island parsing.
\newblock {\em 2011 26th IEEE/ACM International Conference on Automated
  Software Engineering (ASE 2011)}, pages 476--479, 2011.

\bibitem{Ren2020APIMisuseDD}
Xiaoxue Ren, Xinyuan Ye, Zhenchang Xing, Xin Xia, Xiwei Xu, Liming Zhu, and
  Jianling Sun.
\newblock Api-misuse detection driven by fine-grained api-constraint knowledge
  graph.
\newblock {\em 2020 35th IEEE/ACM International Conference on Automated
  Software Engineering (ASE)}, pages 461--472, 2020.

\bibitem{Ye2016LearningTE}
Deheng Ye, Zhenchang Xing, Chee~Yong Foo, J.~Li, and Nachiket Kapre.
\newblock Learning to extract api mentions from informal natural language
  discussions.
\newblock {\em 2016 IEEE International Conference on Software Maintenance and
  Evolution (ICSME)}, pages 389--399, 2016.

\bibitem{Huo2022ARCLINAA}
Yintong Huo, Yuxin Su, Hongming Zhang, and Michael~R. Lyu.
\newblock Arclin: Automated api mention resolution for unformatted texts.
\newblock {\em 2022 IEEE/ACM 44th International Conference on Software
  Engineering (ICSE)}, pages 138--149, 2022.

\bibitem{Li2018ImprovingAC}
Hongwei Li, Sirui Li, Jiamou Sun, Zhenchang Xing, Xin Peng, Mingwei Liu, and
  Xuejiao Zhao.
\newblock Improving api caveats accessibility by mining api caveats knowledge
  graph.
\newblock {\em 2018 IEEE International Conference on Software Maintenance and
  Evolution (ICSME)}, pages 183--193, 2018.

\bibitem{Chen2017UnsupervisedSM}
Chunyang Chen, Zhenchang Xing, and Ximing Wang.
\newblock Unsupervised software-specific morphological forms inference from
  informal discussions.
\newblock {\em 2017 IEEE/ACM 39th International Conference on Software
  Engineering (ICSE)}, pages 450--461, 2017.

\bibitem{Petroni2019LanguageMA}
Fabio Petroni, Tim Rockt{\"a}schel, Patrick Lewis, Anton Bakhtin, Yuxiang Wu,
  Alexander~H. Miller, and Sebastian Riedel.
\newblock Language models as knowledge bases?
\newblock {\em ArXiv}, abs/1909.01066, 2019.

\bibitem{Qiu2020PretrainedMF}
Xipeng Qiu, Tianxiang Sun, Yige Xu, Yunfan Shao, Ning Dai, and Xuanjing Huang.
\newblock Pre-trained models for natural language processing: A survey.
\newblock {\em ArXiv}, abs/2003.08271, 2020.

\bibitem{ma2019easy}
Suyu Ma, Zhenchang Xing, Chunyang Chen, Cheng Chen, Lizhen Qu, and Guoqiang Li.
\newblock Easy-to-deploy api extraction by multi-level feature embedding and
  transfer learning.
\newblock {\em IEEE Transactions on Software Engineering}, 47(10):2296--2311,
  2019.

\bibitem{Zhong2021AFE}
Zexuan Zhong and Danqi Chen.
\newblock A frustratingly easy approach for entity and relation extraction.
\newblock {\em ArXiv}, abs/2010.12812, 2021.

\bibitem{Lu2022UnifiedSG}
Yaojie Lu, Qing Liu, Dai Dai, Xinyan Xiao, Hongyu Lin, Xianpei Han, Le~Sun, and
  Hua Wu.
\newblock Unified structure generation for universal information extraction.
\newblock In {\em ACL}, 2022.

\bibitem{Raffel2020ExploringTL}
Colin Raffel, Noam~M. Shazeer, Adam Roberts, Katherine Lee, Sharan Narang,
  Michael Matena, Yanqi Zhou, Wei Li, and Peter~J. Liu.
\newblock Exploring the limits of transfer learning with a unified text-to-text
  transformer.
\newblock {\em ArXiv}, abs/1910.10683, 2020.

\bibitem{Levine2022StandingOT}
Yoav Levine, Itay Dalmedigos, Ori Ram, Yoel Zeldes, Daniel Jannai, Dor Muhlgay,
  Yoni Osin, Opher Lieber, Barak Lenz, Shai Shalev-Shwartz, Amnon Shashua,
  Kevin Leyton-Brown, and Yoav Shoham.
\newblock Standing on the shoulders of giant frozen language models.
\newblock {\em ArXiv}, abs/2204.10019, 2022.

\bibitem{ye2016software}
Deheng Ye, Zhenchang Xing, Chee~Yong Foo, Zi~Qun Ang, Jing Li, and Nachiket
  Kapre.
\newblock Software-specific named entity recognition in software engineering
  social content.
\newblock In {\em 2016 IEEE 23rd international conference on software analysis,
  evolution, and reengineering (SANER)}, volume~1, pages 90--101. IEEE, 2016.

\bibitem{Devlin2019BERTPO}
Jacob Devlin, Ming-Wei Chang, Kenton Lee, and Kristina Toutanova.
\newblock Bert: Pre-training of deep bidirectional transformers for language
  understanding.
\newblock {\em ArXiv}, abs/1810.04805, 2019.

\bibitem{Landis1977AnAO}
J.~R. Landis and G.~Koch.
\newblock An application of hierarchical kappa-type statistics in the
  assessment of majority agreement among multiple observers.
\newblock {\em Biometrics}, 33 2:363--74, 1977.

\bibitem{Liu2022howto}
Mingwei Liu, Xin Peng, Andrian Marcus, Christoph Treude, Jiazhan Xie, Huanjun
  Xu, and Yanjun Yang.
\newblock How to formulate specific how-to questions in software development?
\newblock In {\em FSE}, 2022.

\bibitem{Kim2014ConvolutionalNN}
Yoon Kim.
\newblock Convolutional neural networks for sentence classification.
\newblock In {\em EMNLP}, 2014.

\bibitem{Joulin2017BagOT}
Armand Joulin, Edouard Grave, Piotr Bojanowski, and Tomas Mikolov.
\newblock Bag of tricks for efficient text classification.
\newblock In {\em EACL}, 2017.

\bibitem{Jiang2021ExploringLE}
Kelvin Jiang, Ronak Pradeep, Jimmy~J. Lin, and David~R. Cheriton.
\newblock Exploring listwise evidence reasoning with t5 for fact verification.
\newblock In {\em ACL}, 2021.

\bibitem{Carmo2020PTT5PA}
Diedre Carmo, Marcos Piau, Israel Campiotti, Rodrigo Nogueira, and Roberto
  de~Alencar~Lotufo.
\newblock Ptt5: Pretraining and validating the t5 model on brazilian portuguese
  data.
\newblock {\em ArXiv}, abs/2008.09144, 2020.

\bibitem{Lewis2020BARTDS}
Mike Lewis, Yinhan Liu, Naman Goyal, Marjan Ghazvininejad, Abdelrahman Mohamed,
  Omer Levy, Veselin Stoyanov, and Luke Zettlemoyer.
\newblock Bart: Denoising sequence-to-sequence pre-training for natural
  language generation, translation, and comprehension.
\newblock In {\em ACL}, 2020.

\bibitem{Brown2020LanguageMA}
Tom~B. Brown, Benjamin Mann, Nick Ryder, Melanie Subbiah, Jared Kaplan,
  Prafulla Dhariwal, Arvind Neelakantan, Pranav Shyam, Girish Sastry, Amanda
  Askell, Sandhini Agarwal, Ariel Herbert-Voss, Gretchen Krueger, T.~J.
  Henighan, Rewon Child, Aditya Ramesh, Daniel~M. Ziegler, Jeff Wu, Clemens
  Winter, Christopher Hesse, Mark Chen, Eric Sigler, Mateusz Litwin, Scott
  Gray, Benjamin Chess, Jack Clark, Christopher Berner, Sam McCandlish, Alec
  Radford, Ilya Sutskever, and Dario Amodei.
\newblock Language models are few-shot learners.
\newblock {\em ArXiv}, abs/2005.14165, 2020.

\bibitem{yin2021api}
Hang Yin, Yuanhao Zheng, Yanchun Sun, and Gang Huang.
\newblock An api learning service for inexperienced developers based on api
  knowledge graph.
\newblock In {\em 2021 IEEE International Conference on Web Services (ICWS)},
  pages 251--261. IEEE, 2021.

\bibitem{Bacchelli2009BenchmarkingLT}
Alberto Bacchelli, Marco D'Ambros, Michele Lanza, and Romain Robbes.
\newblock Benchmarking lightweight techniques to link e-mails and source code.
\newblock {\em 2009 16th Working Conference on Reverse Engineering}, pages
  205--214, 2009.

\bibitem{Li2020AUM}
Xiaoya Li, Jingrong Feng, Yuxian Meng, Qinghong Han, Fei Wu, and Jiwei Li.
\newblock A unified mrc framework for named entity recognition.
\newblock {\em ArXiv}, abs/1910.11476, 2020.

\bibitem{Yan2021NamedER}
Rongen Yan, Xue Jiang, and Depeng Dang.
\newblock Named entity recognition by using xlnet-bilstm-crf.
\newblock {\em Neural Process. Lett.}, 53:3339--3356, 2021.

\bibitem{Yang2019XLNetGA}
Zhilin Yang, Zihang Dai, Yiming Yang, Jaime~G. Carbonell, Ruslan Salakhutdinov,
  and Quoc~V. Le.
\newblock Xlnet: Generalized autoregressive pretraining for language
  understanding.
\newblock In {\em NeurIPS}, 2019.

\bibitem{Zhang2020OpenUEAO}
Ningyu Zhang, Shumin Deng, Zhen Bi, Haiyang Yu, Jiacheng Yang, Mosha Chen, Fei
  Huang, Wei Zhang, and Huajun Chen.
\newblock Openue: An open toolkit of universal extraction from text.
\newblock In {\em EMNLP}, 2020.

\end{thebibliography}

\clearpage

\begin{minipage}[h]{0.25\linewidth}
\includegraphics[height=3.0in,width=1in,clip,keepaspectratio]{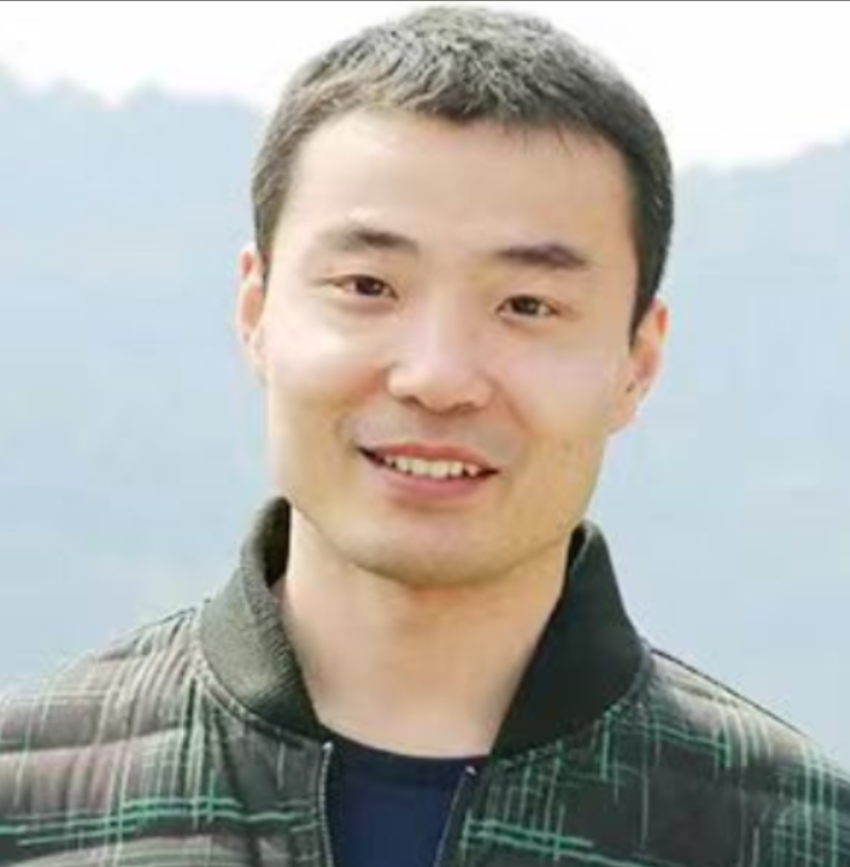}
\end{minipage}
\hfill
\begin{minipage}[h]{0.75\linewidth}
\noindent{\bf QING HUANG}\ 
received the M.S degree in computer application and technology from Nanchang University, in 2009, and the PH.D. degree in computer software and theory from Wuhan University, in 2018. He is currently an Assistant Professor with the School of Computer and Information Engineering, Jiangxi Normal University, China. His research interests include information security, software engineering and knowledge graph.
\end{minipage}

\begin{minipage}[h]{0.25\linewidth}
\includegraphics[height=3.0in,width=1in,clip,keepaspectratio]{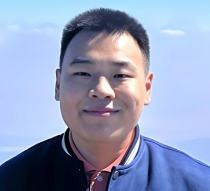}
\end{minipage}
\hfill
\begin{minipage}[h]{0.75\linewidth}
\noindent {\bf Yanbang Sun}\
is a second-year master student at the School of Computer and Information Engineering, Jiangxi Normal University, China. His research interests include software engineering and knowledge graph.
\end{minipage}

\begin{minipage}[h]{0.25\linewidth}
\includegraphics[height=3.0in,width=1in,clip,keepaspectratio]{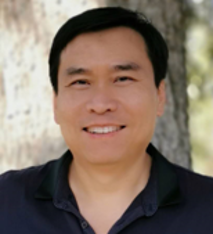}
\end{minipage}
\hfill
\begin{minipage}[h]{0.75\linewidth}
\noindent {\bf Zhenchang Xing}\
is a Senior Research Scientist with Data61, CSIRO, Eveleigh, NSW, Australia. In addition, he is an Associate Professor in the Research School of Computer Science, Australian National University. Previously, he was an Assistant Professor in the School of Computer Science and Engineering, Nanyang Technological University, Singapore, from 2012-2016. His main research areas are software engineering, applied data analytics, and human-computer interaction.
\end{minipage}

\begin{minipage}[h]{0.25\linewidth}
\includegraphics[height=3.0in,width=1in,clip,keepaspectratio]{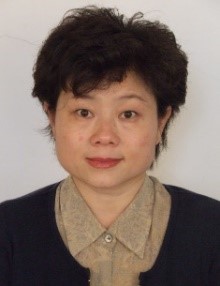}
\end{minipage}
\hfill
\begin{minipage}[h]{0.75\linewidth}
\noindent {\bf MIN YU}\
is a Professor in Communication, Electronic Engineering, and Computer Science at Jiangxi Normal University, was a visiting scholar at the University of California, Irvine, the USA, and interested in Distributed computing, Wireless Sensor Network,  and  Indoor Positioning.
\end{minipage}

\begin{minipage}[h]{0.25\linewidth}
\includegraphics[height=3.0in,width=1in,clip,keepaspectratio]{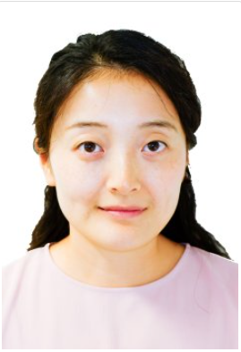}
\end{minipage}
\hfill
\begin{minipage}[h]{0.75\linewidth}
\noindent {\bf Xiwei Xu}\
is a Senior Research Scientist with Architecture\& Analytics Platforms Team, Data61, CSIRO. 
She is also a Conjoint Lecturer with UNSW. 
She started working on blockchain since 2015. 
Her main research interest is software architecture. She also does research in the areas of service computing, business process, and cloud computing and dependability.
\end{minipage}

\begin{minipage}[h]{0.25\linewidth}
\includegraphics[height=3.0in,width=1in,clip,keepaspectratio]{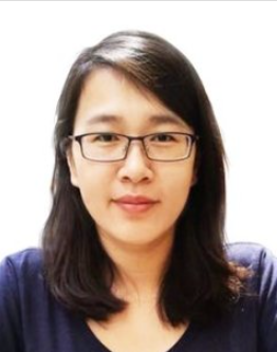}
\end{minipage}
\hfill
\begin{minipage}[h]{0.75\linewidth}
\noindent {\bf Qinghua Lu}\
is a Senior Research Scientist with Data61, CSIRO, Eveleigh, NSW, Australia.  
She has published more than 100 academic papers in international journals and conferences.
Her research interests include the software architecture, blockchain, software engineering for AI, and AI ethics.
\end{minipage}

\end{document}